\documentclass[11pt,letterpaper]{article}
\pdfoutput=1
\usepackage{jheppub}
\usepackage{tikz}
\usepackage{todonotes}

\newcommand{\calQ}{{\cal Q}}

\DeclareMathOperator{\gcr}{cr}
\DeclareMathOperator{\im}{Im}
\DeclareMathOperator{\Li}{Li}
\DeclareMathOperator{\li}{Li}

\DeclareMathOperator{\per}{per}
\DeclareMathOperator{\re}{Re}
\DeclareMathOperator{\vol}{Vol}

\title{One-loop Integrals from Volumes of Orthoschemes}

\author[a]{Lecheng Ren,}
\author[a,b]{Marcus Spradlin,}
\author[c]{Cristian Vergu}
\author[a]{and Anastasia Volovich}

\affiliation[a]{Department of Physics, Brown University,\\
	182 Hope Street, Providence, RI 02912, U.S.A.}
\affiliation[b]{Brown Theoretical Physics Center, Brown University,\\
	340 Brook Street, Providence, RI 02912, U.S.A.}
\affiliation[c]{Niels Bohr International Academy and Discovery Center, Niels Bohr Institute,\\
	University of Copenhagen, Blegdamsvej 17, DK-2100, Copenhagen \O, Denmark}

\emailAdd{lecheng\_ren@brown.edu}
\emailAdd{marcus\_spradlin@brown.edu}
\emailAdd{c.vergu@nbi.ku.dk}
\emailAdd{anastasia\_volovich@brown.edu}

\abstract{Recently in \href{https://arxiv.org/abs/2012.05599}{arXiv:2012.05599} Rudenko presented a formula for the volume of hyperbolic orthoschemes in terms of alternating polylogarithms. We use this result to provide an explicit analytic result for the one-loop scalar $n$-gon Feynman integral in $n$ dimensions, for even $n$, with massless or massive internal and external edges. Furthermore, we evaluate the general six-dimensional hexagon integral in terms of classical polylogarithms.}

\begin{document}
\maketitle

\section{Introduction}

It is well known that Feynman integrals exhibit remarkable mathematical properties. Even the humblest one-loop integrals, which are not only the most relevant for collider physics computations but also feed their structure into higher loop order through differential equations, still have mysteries awaiting discovery. In this paper, we focus on a particularly famous class of one-loop integrals: the scalar $n$-gons in $n$ dimensions. These integrals have seen so much attention that it may be surprising to hear that there is yet more to say.

The box integral has a history stretching back over 60 years; see in particular~\cite{actwu:1961,tHooft:1978jhc,Denner:1991qq,Hodges:2010kq}.  The (five-dimensional) pentagon, in contrast, has been evaluated only much more recently~\cite{Nandan:2013ip,Bourjaily:2019exo}. One key and beautiful mathematical fact about the $n$-gon is that the same integral computes the volume of a simplex in $(n{-}1)$-dimensional hyperbolic space whose shape is set by the (Euclidean) kinematic data and propagator masses~\cite{Davydychev:1997wa,Mason:2010pg}. The box integral was revisited in this context in~\cite{Bourjaily:2019exo}, making use of the Murakami-Yano formula for the volume of a hyperbolic tetrahedron~\cite{MR2154824}. Other papers that have made use of this correspondence include~\cite{Schnetz:2010pd,Paulos:2012qa,Arkani-Hamed:2017ahv,Davydychev:2017bbl}, and the simplicity of the $n$-gon in Mellin space was pointed out in~\cite{Paulos:2012nu}.

Around a decade ago there was a brief flurry of interest in the hexagon integral~\cite{DelDuca:2011ne,Dixon:2011ng,DelDuca:2011jm,DelDuca:2011wh,Spradlin:2011wp}, due in part to the fact that (for special kinematics) it is related by differential equations to some one- and two-loop integrals that appear in planar $\mathcal{N}=4$ super-Yang Mills theory. Also, the time was ripe for progress on loop integrals since the symbol calculus for polylogarithm functions had recently been introduced to the physics literature~\cite{Goncharov:2010jf}. However, while its symbol was certainly known~\cite{Spradlin:2011wp}, the problem of explicitly evaluating the fully general hexagon integral remained unsolved, and seemed exceedingly difficult since the symbol involves 16 distinct square roots of 9 independent kinematic cross-ratios. Today the hexagon integral continues to play a starring role in developments at the forefront of multi-loop integration, in part because it is related by a simple differential equation~\cite{Paulos:2012nu} to the elliptic polylogarithmic double box integral which has received considerable recent attention~\cite{Bourjaily:2017bsb,Adams:2018bsn,Adams:2018kez,Ananthanarayan:2020ncn,Bloch:2021hzs,Kristensson:2021ani,Wilhelm:2022wow,Pozo:2022dox}. The (IR-divergent) massless hexagon has been evaluated to ${\mathcal{O}}(\epsilon)$ in dimensional regularization in~\cite{Henn:2022ydo}.

In this paper, we provide\footnote{Not only in the sense of formulas in the text, but perhaps of more practical importance, also in an ancillary Mathematica file.}
an explicit analytic formula for the hexagon integral with arbitrary internal and external masses, for Euclidean kinematics, in terms of the functions $\log$, $\Li_2$ and $\Li_3$ with algebraic functions of cross-ratios as their arguments.  Our work was made possible by a recent mathematical breakthrough due to Rudenko, building on the classic work of Coxeter and B{\"o}hm~\cite{Cox35,Cox36,Boh60,Boh64}, who provided an explicit formula for the volume of any (odd-dimensional) hyperbolic orthoscheme in terms of a class of functions called alternating polylogarithms~\cite{Rudenko}. Orthoschemes are generalizations of right triangles and we can compute the volume of any simplex---and hence any (even) $n$-gon integral---by recursively expressing the volume of the simplex as a sum\footnote{As we explain in Sec.~\ref{sec:dissection}, this is not a dissection since some components may lie outside the original simplex and therefore contribute to the volume with a minus sign. It is a conjecture of Hadwiger that any simplex can be dissected into orthoschemes.} of volumes of orthoschemes.\footnote{For odd $n$ the $n$-gon integral can be represented as a linear combination of $m$-gon integrals with even $m$ and $m<n$ via a generalized version of the Gauss-Bonnet theorem~\cite{Bernig, Schnetz:2010pd}.} Rudenko's construction works for any even $n$, but $n=2, 4, 6$ are special because only in these cases can the answer be expressed in terms of the classical polylogarithm functions\footnote{Interestingly the 7- and 9-gon integrals (with massless propagators) should also be classical; we leave these as exercises for the enthusiastic reader.}, and we have invested considerable effort in doing so. (The main functions provided in the ancillary file work for any even $n$, limited only by computational power.)

Scientific progress is often nonlinear, and it was only during the course of our work that we learned of a 1995 paper in which Kellerhals provided an explicit formula (as a one-fold integral of weight-2 polylogarithms) for the volume of a doubly-asymptotic hyperbolic orthoscheme in 5 dimensions~\cite{Kellerhals}. Doubly-asymptotic means having at least two vertices on the boundary, so combined with our triangulation algorithm, this is enough to compute the hexagon integral for massless internal propagators but still arbitrary external kinematics---a paper that should have been written a decade ago during the hexagon's brief heyday.

Before jumping into the details of our calculation let us make two remarks. First, it is interesting to note that the so-called antipodal map, which reverses the order of the entries in a symbol and has made an interesting appearance in the recent literature~\cite{Dixon:2021tdw,Liu:2022vck}, pops up in our work in two seemingly unrelated ways. The symbol of the $n$-gon integral is reversed if one exchanges the Gram matrix $Q$ (see Sec.~\ref{sec:sec2}) associated to a simplex with $Q^{-1}$, and the ``real period'' map~\cite{Goncharov} reviewed in Appendix~\ref{app:realperiod} has the effect of reversing the symbol of a function.

Second, it is worth emphasizing that even after the present work, the story of the $n$-gon integral is certainly not yet fully told.  For one thing, in order to cautiously establish a foothold in uncharted territory we restrict our attention to the safe harbor of Euclidean kinematics, but it would certainly be interesting to better understand the structure of the alternating polylogarithm representation in other regions and kinematic signatures (see~\cite{Bourjaily:2019exo} for a nice discussion, and~\cite{Loebbert:2019vcj,Corcoran:2020epz} for a bootstrap approach to computing integrals where understanding various kinematic regions plays a crucial role). In principle, one has to keep track of the branch cuts that are crossed during analytic continuation to the correct region in kinematic space, and add the proper discontinuities across these branch cuts. Practically, this can be conveniently achieved by ``bootstraping'' the final result by taking a linear combination of our formula with all its discontinuities and matching with the numerical result of the integral. Moreover, although our triangulation algorithm gets the job done, it necessarily mangles some of the $n$-gon integral's structure: in particular, our result is not manifestly cyclically invariant. We hope and anticipate that the full structure of the $n$-gon integral will be better elucidated in the future.

This paper is structured as follows. In Sec.~\ref{sec:sec2} we very briefly review the relation between $n$-gon integrals and the volumes of hyperbolic $(n{-}1)$-simplices. In Sec.~\ref{sec:realperiod} we describe the main ideas behind Rudenko's construction and explain our triangulation algorithm which leads to our main result~(\ref{eq:volsimpsgn}). In Sec.~\ref{sec:kellerhals} we review Kellerhals's formula, and in several appendices we compile various necessary technical details.

\section{\texorpdfstring{The one-loop $n$-gon integral as a hyperbolic volume}{The one-loop n-gon integral as a hyperbolic volume}}
\label{sec:sec2}

We begin by briefly reviewing the well-known correspondence between one-loop Feynman integrals and volumes of hyperbolic simplices~\cite{Davydychev:1997wa,Mason:2010pg} (see also~\cite{Bourjaily:2019exo} for a nice review and many additional details). The one-loop $n$-gon scalar Feynman integral in $n$-dimensional Euclidean space, with propagators having arbitrary masses, is written in terms of dual (or region) momenta as
\begin{equation}
    I_n(x_1, \ldots, x_n) = \int \frac{d^n x}{\pi^{n/2}} \prod_{i=1}^n \frac{1}{(x - x_i)^2 + m_i^2}\,.
\end{equation}
After introducing Feynman parameters and integrating out $x$, this can be expressed as
\begin{equation}
    \label{eq:idef}
    I_n(G_{ij}) = \Gamma\left( \frac{n}{2} \right) \int_0^\infty \prod_{i=1}^n d\alpha_i\, \delta(\alpha_n - 1) \left( \sum_{i,j=1}^n G_{ij} \alpha_i \alpha_j\right)^{-n/2}
\end{equation}
in terms of the quadratic form $G_{ij} = \frac{1}{2}((x_i-x_j)^2 + m_i^2 + m_j^2)$. Henceforth we view $I_n$ as a function of the quadratic form $G$ directly, instead of as a function of the $x_i$ and $m_i$.

Next, we turn our attention to hyperbolic volumes.  Consider the $(n{-}1)$-dimensional hyperboloid defined by the equation
\begin{equation}
    y_1^2 + y_2^2 + \cdots + y_{n-1}^2 - y_n^2 = - 1
\end{equation}
in $n$-dimensional Euclidean space. An alternate representation of this hyperboloid, called the \emph{projective model}, associates to every point
\begin{equation}
    y = (y_1, \ldots, y_{n-1}, \sqrt{1 + y_1^2 + \cdots + y_{n-1}^2})
\end{equation}
on the upper branch of the hyperboloid the point $p$ in the unit $(n{-}1)$-dimensional ball given by
\begin{equation}
    p = (p_1, \ldots, p_{n-1}) \qquad \text{with} \qquad p_i = y_i/\sqrt{1 + y_1^2 + \cdots + y_{n-1}^2}\,.
\end{equation}
Equivalently we could view $y_i = p_i/\sqrt{1 - p_1^2 - \cdots - p_{n-1}^2}$ with the convention $p_n = 1$. We denote the inner product of two points in the projective model by
\begin{equation}
    \label{eq:Qpq}
    \langle p, q\rangle  := 1 - \sum_{i=1}^{n-1} p_i q_i\,.
\end{equation}
For any point $p$ we have $0 \le \langle p,p \rangle \le 1$, with $\langle p,p \rangle = 0$ if and only if $p$ lies on the $S^{n-2}$ boundary of the unit ball.

Now consider a geodesic simplex in the hyperboloid with vertices $v_1, \ldots, v_n$. We define the \emph{Gram matrix} $Q$ associated to the simplex by
\begin{equation}
\label{eq:gramdef}
    Q_{ij} = \langle v_i, v_j \rangle\,.
\end{equation}
The volume of this hyperbolic simplex depends only on the quadratic form $Q$ and can be represented by the integral
\begin{equation}
    \label{eq:voldef}
    \text{Vol}(Q) = \sqrt{|\det Q|} \int_0^\infty \prod_{i=1}^n d\alpha_i\, \delta(\alpha_n-1) \left( \sum_{i,j=1}^n Q_{ij} \alpha_i \alpha_j\right)^{-n/2}\,.
\end{equation}
Note that this is manifestly invariant under rescaling $Q_{ij} \to Q_{ij} \lambda_i \lambda_j$; all calculations below will only involve ratios that are invariant under such scalings. By comparing~(\ref{eq:idef}) to~(\ref{eq:voldef}) we see that the one-loop $n$-point integral in $n$ dimensions can be expressed as
\begin{equation}
    \label{eq:ivolrelation}
    I_n(Q) = \Gamma\left( \frac{n}{2} \right) \frac{\text{Vol}(Q)}{\sqrt{|\det Q|}} \qquad \text{with} \qquad Q_{ij} = \frac{(x_i-x_j)^2 + m_i^2 + m_j^2}{2}\,.
\end{equation}
Note that the $Q_{ij}$ appearing here is not the Gram matrix of any simplex computed from~(\ref{eq:gramdef}), but can be brought to such a form by an appropriate rescaling. However, the presentation~(\ref{eq:ivolrelation}) has the virtue of making the massless limit more transparent: the diagonal entries are now $Q_{ii} = m_i^2$, so an integral with all massless propagators computes the volume of a simplex with all vertices on the boundary, which is called an \emph{ideal} simplex.

\section{Volumes of hyperbolic simplices}
\label{sec:realperiod}

In order to compute the volume of the hyperbolic simplex appearing in~(\ref{eq:ivolrelation}) we proceed in two steps. First, we review the recent mathematical breakthrough~\cite{Rudenko} that expresses the volume of a special class of simplices, called \emph{hyperbolic orthoschemes}, in terms of alternating multiple polylogarithms. Then in Sec.~\ref{sec:dissection} we explain a recursive procedure for dividing any simplex into orthoschemes, thereby allowing us to compute~(\ref{eq:ivolrelation}) for any $Q$. In this section, we use $Q$ ($\calQ$) interchangeably to refer either to a simplex (orthoscheme) or to its associated Gram matrix.

\subsection{Volumes of orthoschemes}
\label{sec:realperiods}

An $(n{-}1)$-dimensional hyperbolic orthoscheme is a hyperbolic simplex for which the bounding hyperplanes $H_i$ can be ordered $(H_1, \ldots, H_n)$ in such a way that $H_i$ is orthogonal to $H_j$ for $|i-j|>1$. An equivalent definition is that there exists an ordering of the vertices $(v_1,\ldots,v_n)$ of the orthoscheme so that the edges $(v_1,v_2)$, $(v_2,v_3)$, $\ldots$, $(v_{n-1},v_n)$ are mutually orthogonal. The equivalence of these two definitions can be seen by taking the hyperplane $H_i$ to be the convex hull of all vertices except $v_i$.

A \emph{subscheme} of an orthoscheme is the convex hull of an arbitrary subset of its vertices, denoted by $\langle I \rangle = \text{conv} \{ v_i |\, i \in I \}$ with $I$ being an ordered subset of $\{1,\ldots, n\}$. One can show that any subscheme of an orthoscheme is again an orthoscheme.

It was shown in~\cite{Cox36} that there is a bijection between $(n{-}1)$-dimensional orthoschemes and configurations $z = (z_0, z_1, \ldots, z_{n+1}) \in \mathcal{M}_{0,n+2}$, the moduli space of $n{+}2$ points in $\mathbb{P}^1$. Under this bijection the Gram matrix of the orthoscheme is expressed as~\cite[(6.4), Example 6.6]{Rudenko}
\begin{equation}
\label{eq:moduli2orthos}
    \calQ_{i,j} = \mathcal{Q}_{j,i} = \lambda_i \lambda_j (z_0-z_i) (z_j-z_{n+1}) \,, \qquad 1 \le i \leq j \le n \,,
\end{equation}
where the $\lambda_i$ are free parameters (associated to rescaling the Feynman parameters, as discussed in the previous section). The condition for the orthoscheme to be hyperbolic is
\begin{align}
\label{eq:hyperbolic}
    \calQ_{i,j}^2 > \calQ_{i,i} \calQ_{j,j}
\end{align}
for each $i\ne j$, which we henceforth assume. (This will always be true when we study the $n$-gon integral in generic Euclidean kinematics~\cite{Bourjaily:2019exo}.) In addition one can immediately see that~(\ref{eq:moduli2orthos}) satisfies
\begin{equation}
    \label{eq:pythag}
    \calQ_{i,j} \calQ_{j,k} = \calQ_{i,k} \calQ_{j,j}\,, \qquad i<j<k \,,
\end{equation}
which is the hyperbolic Pythagorean theorem.

Inverting the relation~(\ref{eq:moduli2orthos}) allows one to represent every cross-ratio of the configuration $z \in \mathcal{M}_{0,n+2}$ in terms of the Gram matrix $\calQ$ of the corresponding hyperbolic orthoscheme:
\begin{equation}
    \label{eq:orthos2moduli}
    \begin{aligned}
        & \frac{z_{0,a} z_{b,n+1}}{z_{0,n+1} z_{b,a}} = \frac{\calQ_{a,b}^2}{\calQ_{a,b}^2 - \calQ_{a,a} \calQ_{b,b}}\,,\\
        & \frac{z_{0,a} z_{b,c}}{z_{0,c} z_{b,a}} = -\frac{\calQ_{a,b}^2 (\calQ_{b,c}^2 - \calQ_{b,b}\calQ_{c,c})}{(\calQ_{a,b}^2 - \calQ_{a,a}\calQ_{b,b}) \calQ_{b,b} \calQ_{c,c}}\,,\\
        & \frac{z_{a,b} z_{c,n+1}}{z_{a,n+1} z_{c,b}} = -\frac{(\calQ_{a,b}^2 - \calQ_{a,a}\calQ_{b,b}) \calQ_{b,c}^2}{\calQ_{a,a} \calQ_{b,b} (\calQ_{b,c}^2 - \calQ_{b,b}\calQ_{c,c})}\,,\\
        & \frac{z_{a,b} z_{c,d}}{z_{a,d} z_{c,b}} = -\frac{\calQ_{b,c}^2 (\calQ_{a,b}^2 - \calQ_{a,a}\calQ_{b,b}) (\calQ_{c,d}^2 - \calQ_{c,c}\calQ_{d,d})}{\calQ_{b,b} \calQ_{c,c} (\calQ_{a,d}^2 - \calQ_{a,a}\calQ_{d,d}) (\calQ_{b,c}^2 - \calQ_{b,b}\calQ_{c,c})}
    \end{aligned}
\end{equation}
for $1\le a<b<c<d\le n$, where $z_{i,j} = z_i - z_j$. After fixing the convenient gauge
\begin{equation}
\label{eq:gauge}
    z_0 = 0\,, \qquad z_n = 1\,, \qquad z_{n+1} = \infty
\end{equation}
it follows from~(\ref{eq:orthos2moduli}) that the remaining $z_i$ can be expressed simply as
\begin{equation}
    \label{eq:zidef}
    z_i = \frac{\calQ_{i,n}^2}{\calQ_{i,i} \calQ_{n,n}}\,.
\end{equation}
For our application to the $n$-gon integral in generic Euclidean kinematics, the $z_i$ will always be ordered as $z_0 < z_n < z_{n-1} < \cdots < z_1 < z_{n+1}$, opposite (but equivalent) to the convention in~(6.12) of~\cite{Rudenko}. The formulas~(\ref{eq:orthos2moduli}) and~(\ref{eq:zidef}) break down when a vertex of the orthoscheme is on the boundary, a situation that we will encounter unless all of the propagators in the integral are massive. This indicates only an incompatibility with our gauge choice~(\ref{eq:gauge}), not a breakdown of Rudenko's formula. In practice, we sidestep this problem by adding a mass $\epsilon$ to each propagator and then taking the limit $\epsilon\to 0$, under which everything is smooth for generic Euclidean kinematics, at the end of the calculation. (This limit can be taken analytically, and should therefore be thought of more as a bookkeeping trick; it does not require numerical extrapolation.)

Rudenko showed that the volume of any odd-dimensional orthoscheme can be expressed in terms of these cross-ratios on the corresponding moduli space. The explicit construction in~\cite{Rudenko} is rather involved, so we defer the details to Appendix~\ref{sec:rudenkodetails} and simply quote here the results for the volume of a 3-dimensional orthoscheme relevant for the four-dimensional box integral (here and in the following we write $z_{i,j}$ as $z_{ij}$ when no confusion can arise):
\begin{equation}
    \label{eq:vol3ort}
    \begin{aligned}
        \text{Vol}(\calQ_{n=4}) = \frac{1}{2} \per\left\lbrack \text{ALi}_{1,1}\left(-\frac{z_{01} z_{25}}{z_{05} z_{12}},-\frac{z_{23} z_{45}}{z_{25} z_{34}}\right) \right. &\left. -\,\text{ALi}_{1,1}\left(-\frac{z_{01} z_{45}}{z_{05} z_{14}},-\frac{z_{14} z_{23}}{z_{12} z_{34}}\right)\right.\\
        &\left. +\, \text{ALi}_{1,1}\left(-\frac{z_{03} z_{45}}{z_{05} z_{34}},-\frac{z_{01} z_{23}}{z_{03} z_{12}}\right) \right\rbrack,
    \end{aligned}
\end{equation}
and for a 5-dimensional orthoscheme relevant for the six-dimensional hexagon integral:
\begin{equation}
    \label{eq:vol5ort}
    \begin{aligned}
        \text{Vol}(\calQ_{n=6}) =& \frac{1}{8} \per \bigg\lbrack\, \text{ALi}_{1,2}\left(-\frac{z_{01} z_{67}}{z_{07} z_{16}},\frac{z_{16} z_{23} z_{45}}{z_{12} z_{34} z_{56}}\right) - \text{ALi}_{1,2}\left(-\frac{z_{01} z_{47}}{z_{07} z_{14}},\frac{z_{14} z_{23} z_{45} z_{67}}{z_{12} z_{34} z_{47} z_{56}}\right)\\
        & + \text{ALi}_{1,2}\left(-\frac{z_{03} z_{47}}{z_{07} z_{34}},\frac{z_{01} z_{23} z_{45} z_{67}}{z_{03} z_{12} z_{47} z_{56}}\right)-\text{ALi}_{1,2}\left(-\frac{z_{03} z_{67}}{z_{07} z_{36}},\frac{z_{01} z_{23} z_{36} z_{45}}{z_{03} z_{12} z_{34} z_{56}}\right)\\
        - \text{ALi}_{1,1,1}&\left(-\frac{z_{01} z_{27}}{z_{07} z_{12}},-\frac{z_{23} z_{47}}{z_{27} z_{34}},-\frac{z_{45} z_{67}}{z_{47} z_{56}}\right)+\text{ALi}_{1,1,1}\left(-\frac{z_{01} z_{27}}{z_{07} z_{12}},-\frac{z_{23} z_{67}}{z_{27} z_{36}},-\frac{z_{36} z_{45}}{z_{34} z_{56}}\right)\\
        - \text{ALi}_{1,1,1}&\left(-\frac{z_{01} z_{27}}{z_{07} z_{12}},-\frac{z_{25} z_{67}}{z_{27} z_{56}},-\frac{z_{23} z_{45}}{z_{25} z_{34}}\right)+\text{ALi}_{1,1,1}\left(-\frac{z_{01} z_{47}}{z_{07} z_{14}},-\frac{z_{14} z_{23}}{z_{12} z_{34}},-\frac{z_{45} z_{67}}{z_{47} z_{56}}\right)\\
        + \text{ALi}_{1,1,1}&\left(-\frac{z_{01} z_{47}}{z_{07} z_{14}},-\frac{z_{45} z_{67}}{z_{47} z_{56}},-\frac{z_{14} z_{23}}{z_{12} z_{34}}\right)-\text{ALi}_{1,1,1}\left(-\frac{z_{01} z_{67}}{z_{07} z_{16}},-\frac{z_{16} z_{23}}{z_{12} z_{36}},-\frac{z_{36} z_{45}}{z_{34} z_{56}}\right)\\
        + \text{ALi}_{1,1,1}&\left(-\frac{z_{01} z_{67}}{z_{07} z_{16}},-\frac{z_{16} z_{25}}{z_{12} z_{56}},-\frac{z_{23} z_{45}}{z_{25} z_{34}}\right)-\text{ALi}_{1,1,1}\left(-\frac{z_{01} z_{67}}{z_{07} z_{16}},-\frac{z_{16} z_{45}}{z_{14} z_{56}},-\frac{z_{14} z_{23}}{z_{12} z_{34}}\right)\\
        - \text{ALi}_{1,1,1}&\left(-\frac{z_{03} z_{47}}{z_{07} z_{34}},-\frac{z_{01} z_{23}}{z_{03} z_{12}},-\frac{z_{45} z_{67}}{z_{47} z_{56}}\right)-\text{ALi}_{1,1,1}\left(-\frac{z_{03} z_{47}}{z_{07} z_{34}},-\frac{z_{45} z_{67}}{z_{47} z_{56}},-\frac{z_{01} z_{23}}{z_{03} z_{12}}\right)\\
        + \text{ALi}_{1,1,1}&\left(-\frac{z_{03} z_{67}}{z_{07} z_{36}},-\frac{z_{01} z_{23}}{z_{03} z_{12}},-\frac{z_{36} z_{45}}{z_{34} z_{56}}\right)+\text{ALi}_{1,1,1}\left(-\frac{z_{03} z_{67}}{z_{07} z_{36}},-\frac{z_{36} z_{45}}{z_{34} z_{56}},-\frac{z_{01} z_{23}}{z_{03} z_{12}}\right)\\
        - \text{ALi}_{1,1,1}&\left(-\frac{z_{05} z_{67}}{z_{07} z_{56}},-\frac{z_{01} z_{25}}{z_{05} z_{12}},-\frac{z_{23} z_{45}}{z_{25} z_{34}}\right)+\text{ALi}_{1,1,1}\left(-\frac{z_{05} z_{67}}{z_{07} z_{56}},-\frac{z_{01} z_{45}}{z_{05} z_{14}},-\frac{z_{14} z_{23}}{z_{12} z_{34}}\right)\\
        - \text{ALi}_{1,1,1}&\left(-\frac{z_{05} z_{67}}{z_{07} z_{56}},-\frac{z_{03} z_{45}}{z_{05} z_{34}},-\frac{z_{01} z_{23}}{z_{03} z_{12}}\right)\bigg\rbrack.
    \end{aligned}
\end{equation}
These are expressed in terms of the alternating multiple polylogarithm defined by
\begin{equation}
\label{eq:ALidef}
    \text{ALi}_{m_1,\ldots, m_k}( \varphi_1, \ldots, \varphi_k ) := \sum_{\epsilon_1, \ldots, \epsilon_k \in \{-1,1\}} \left(\prod_{i=1}^{k} \frac{\epsilon_i}{2}\right) \text{Li}_{m_1, \ldots, m_k} (\epsilon_1 \sqrt{\varphi_1}, \ldots, \epsilon_k \sqrt{\varphi_k}) \,,
\end{equation}
and the function ``$\text{per}$'' appearing above is the \emph{real period} reviewed in Appendix~\ref{app:realperiod}; there we also provide explicit formulas for the real periods of all multiple polylogarithms of weight up to 3.

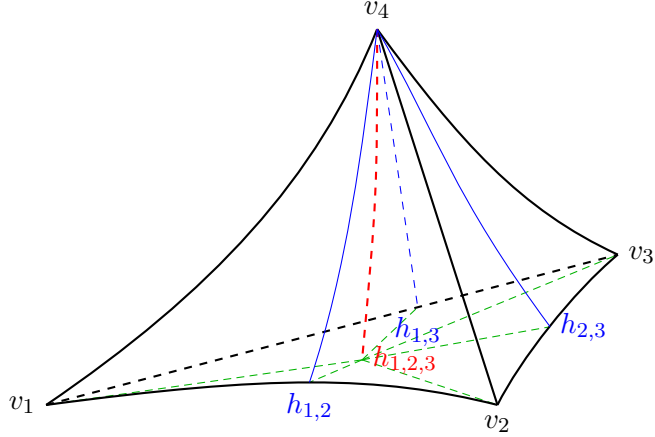
\begin{figure}
    \centering
    \begin{tikzpicture}[scale=2]
        \coordinate (v1) at (-3,-1);
        \coordinate (v2) at (0,-1);
        \coordinate (v3) at (0.8,0);
        \coordinate (v4) at (-0.8,1.5);
        \coordinate (h12) at (-1.25,-0.85);
        \coordinate (h23) at (0.35,-0.48);
        \coordinate (h13) at ({-3*0.35+0.8*0.65},{-1*0.35});
        \coordinate (h123) at (-0.9,-0.7);

        \draw[densely dashed,black!30!green] (h123) -- (h12) (h123) -- (h23) (h123) -- (h13);
        \draw[densely dashed,black!30!green] (h123) -- (v1) (h123) -- (v2) (h123) -- (v3);

        \draw[thick] (v1) .. controls (-1.5,-0.8) and (-0.8,-0.8) .. (v2) node[below]{$v_2$};
        \draw[thick] (v2) .. controls (0.15,-0.7) and (0.5,-0.25) .. (v3) node[right]{$v_3$};
        \draw[thick] (v3) .. controls (0.15,0.3) and (-0.25,0.75) .. (v4) node[above]{$v_4$};
        \draw[thick] (v4) .. controls (-1.2,0.5) and (-1.9,-0.25) .. (v1) node[left]{$v_1$};
        \draw[thick] (v4) -- (v2);
        \draw[thick,dashed] (v1) -- (v3);

        \draw[thick,dashed,red] (v4) .. controls (-0.8,0.4) .. (h123) node[right]{$h_{1,2,3}$};

        \draw[blue] (v4) .. controls (-0.9,0.7) and (-1,-0.1) .. (h12) node[below]{$h_{1,2}$};
        \draw[blue] (v4) .. controls (-0.4,0.8) and (-0.3,0.4) .. (h23) node[right]{$h_{2,3}$};
        \draw[blue,dashed] (v4) -- (h13) node[below]{$h_{1,3}$};
    \end{tikzpicture}
    \caption{Triangulation of a hyperbolic tetrahedron into six orthoschemes. The point $h_{I}$ denotes the point where the altitude from $v_4$ intersects the boundary $\langle I\rangle$, defined as the convex hull of vertices $\{ v_i |\, i \in I\}$. The six orthoschemes are $\{ \text{conv}(v_4, h_{1,2,3}, h_{i,j}, v_i) |\, i,j\in\{1,2,3\},\, i\neq j \}.$}
    \label{fig:ort3}
\end{figure}

\subsection{Triangulating a simplex into orthoschemes}
\label{sec:dissection}

In this section we discuss a triangulation of hyperbolic simplices into orthoschemes, beginning with the case of a hyperbolic tetrahedron as an example. The triangulation, shown in Fig.~\ref{fig:ort3}, proceeds as follows. Let the vertices be ($v_1,v_2,v_3,v_4$). First we find the altitude (shown in red) from the vertex $v_4$ to the plane $(v_1, v_2, v_3)$, and then the altitudes (shown in blue) from $v_4$ to the edges $(v_1,v_2)$, $(v_2,v_3)$ and $(v_1,v_3)$. Then the tetrahedron has been divided into six simplices as shown in the figure: the convex hulls of $(v_4,h_{1,2,3},h_{2,3},v_3)$, $(v_4,h_{1,2,3},h_{2,3},v_2)$, $(v_4,h_{1,2,3},h_{1,3},v_1)$, $(v_4,h_{1,2,3},h_{1,3},v_3)$, $(v_4,h_{1,2,3},h_{1,2},v_2)$ and $(v_4,h_{1,2,3},h_{1,2},v_1)$. One can show that these simplices are all orthoschemes.

This algorithm generalizes recursively. Starting with an $(n{-}1)$-dimensional simplex with vertices $(v_1,\ldots,v_n)$, we first draw the altitudes from $v_n$ to each boundary of dimension $\ge 1$. Each boundary is a convex hull of some subset of (at least two of) the vertices $(v_1,\ldots,v_{n-1})$ and we denote these subsimplices by
\begin{equation}
    \langle I \rangle := \text{conv}(\{v_i | \, i\in I\})
\end{equation}
for any ordered subset $I$ of $(v_1,\ldots,v_{n-1})$ of cardinality $2 \leq |I| \leq n-1$. We also denote the altitude from $v_n$ to $\langle I \rangle$ by $h_I$.

Then for every permutation $\sigma$ of $\{1,\ldots,n{-}1\}$ there is an orthoscheme
\begin{multline}
    \label{eq:dislabel}
    \calQ^\sigma := \text{conv} (v_{\sigma(1)}, \, h_{\sigma(1), \sigma(2)}, \, h_{\sigma(1), \sigma(2), \sigma(3)}, \, \ldots, \, h_{\sigma(1), \sigma(2) ,\cdots, \sigma(n-2)},\\
    h_{\sigma(1), \sigma(2) ,\cdots, \sigma(n-1)}=h_{1,2,\cdots,n-1}, v_n)\,.
\end{multline}
These $(n-1)!$ orthoschemes do not always provide a dissection because some components may lie outside the original simplex. Such components contribute to the total volume with a minus sign, which we will account for shortly.

Next we discuss how to represent the cross-ratios associated to each orthoscheme in terms of the Gram matrix $Q$ associated to the initial simplex. For any subset $I$ of $\{1,\ldots,n\}$, we denote by $Q[I]$ the submatrix of $Q$ obtained by keeping the rows and columns indexed by the elements of $I$. Using this notation, the moduli space coordinates $z^\sigma = (z_0^\sigma, \ldots, z_{n+1}^\sigma)$ associated to the orthoscheme $\calQ^\sigma$ are given by
\begin{equation}
\label{eq:simp2moduli}
    z^{\sigma}_i = 1 - \frac{\det Q[\sigma(1),\ldots, \sigma(i), n]}{\det Q[\sigma(1), \ldots,\sigma(i)] \, Q_{n,n}} \,, \qquad1 \le i \le n-1\,,
\end{equation}
with $z_0^\sigma = 0$, $z_n^\sigma = 1$, and $z_{n+1}^{\sigma} = \infty$ for each $\sigma$ (according to our choice of gauge). Then the volume of $\calQ^\sigma$ is given by evaluating Rudenko's formula~(\ref{eq:volinorthos}) on~(\ref{eq:simp2moduli}):
\begin{equation}
   \label{eq:volortingram}
    \text{Vol}(\calQ^{\sigma}) = (\text{eq.~\ref{eq:volinorthos}})\bigg|_{\text{eq.~(\ref{eq:simp2moduli})}} \,.
\end{equation}

As mentioned above, in general it can happen that some orthoschemes fall outside the original simplex and therefore contribute negatively to the volume. This can be accounted for by comparing the sign of the determinant of the Gram matrix of each orthoscheme $\calQ^\sigma$ to that of the original simplex. Altogether, the complete formula for the volume of any odd-dimensional hyperbolic simplex $Q$ may be written as
\begin{equation}
\label{eq:volsimpsgn}
\boxed{
    \text{Vol}(Q) = \sum_{\sigma \in S_{n-1}} \frac{ \text{sgn}(\sigma)\, \text{sgn}(\det \calQ^{\sigma}) }{\text{sgn}(\det Q)}\, (\text{eq.~\ref{eq:volinorthos}}) \bigg|_{\text{eq.~(\ref{eq:simp2moduli})}}
    }
\end{equation}
where $\text{sgn}(\sigma)$ denotes the signature of the permutation $\sigma$. The determinant $\det \calQ^{\sigma}$ can also be computed recursively according to our triangulation algorithm, although the full result is a bit lengthy so we will not write it here. Plugging~(\ref{eq:volsimpsgn}) into~(\ref{eq:ivolrelation}) completes our calculation of the scalar $n$-dimensional $n$-gon integral (for even $n$), for arbitrary internal and external masses, in terms of the Gram matrix $Q$ given in~(\ref{eq:ivolrelation}).

\section{An alternative formula for the hexagon with massless propagators}
\label{sec:kellerhals}

According to the discussion in the previous sections, Rudenko's formula can be used to compute the one-loop scalar $n$-gon Feynman integral in $n$ dimensions for any even $n$. It applies for general Euclidean kinematics, including arbitrary internal and external masses, but is rather complicated to use in practice. For example, for $n=6$ one has to combine~(\ref{eq:vol5ort}), (\ref{eq:simp2moduli}), (\ref{eq:volsimpsgn}), and (\ref{eq:per3})--(\ref{eq:per111}). However, if we take all of the propagators to be massless (but still with arbitrary external momenta), a simpler formula for $n=6$ is available thanks to a 1995 result due to Kellerhals~\cite{Kellerhals}, which we now describe.

A massless propagator in the Feynman integral corresponds to a point on the boundary of the hyperboloid. When all propagators are massless, the corresponding hyperbolic simplex is therefore ideal. When we triangulate a simplex using the algorithm described in Sec.~\ref{sec:dissection}, the first and last vertices of every orthoscheme  are always vertices of the original simplex. Therefore, every orthoscheme appearing in the triangulation of an ideal simplex is \emph{doubly asymptotic}, which means having (at least) two vertices on the boundary.

The volume of doubly asymptotic simplices in 5 dimensions was computed by Kellerhals in~\cite{Kellerhals}. The volume is expressed in terms of the dihedral angles
\begin{equation}
    \alpha_i := \angle (H_i, H_{i+1}) \,, \quad i \in \{1, 2, 3, 4, 5\}\,.
\end{equation}
For a generic orthoscheme the five dihedral angles are independent. However, for a doubly asymptotic orthoscheme, the dihedral angles satisfy the relation
\begin{equation}
    \tan \alpha_1 \cot \alpha_4 = \cot \alpha_2 \tan \alpha_5 = \cot \alpha_0 \tan \alpha_3 = \tan \Theta =: \lambda
\end{equation}
where we have introduced two new angles $\alpha_0$ and $\Theta$ and one more parameter $\lambda$ for later convenience, with $0 \leq \alpha_0 \leq \frac{\pi}{2}$, $0 \leq \Theta \leq \frac{\pi}{2}$, and $\lambda \ge 0$.

Kellerhals's formula for the volume of a double asymptotic 5-dimensional orthoscheme is
\begin{equation}\label{eq:voldoubort}
    \begin{aligned}
        \vol(&\calQ_{n=6}^{\text{da}})=
        -\frac{1}{8}\Big\{I\left(\lambda^{-1}, 0 ; \alpha_1\right)+\frac{1}{2} I\left(\lambda, 0 ; \alpha_2\right)-I\left(\lambda^{-1}, 0 ; \frac{\pi}{2} - \alpha_0 \right)\\
        &\qquad\quad ~+\frac{1}{2}I\left(\lambda, 0 ; \alpha_4\right)+I\left(\lambda^{-1}, 0 ; \alpha_5\right)\Big\}\\
        &+\frac{1}{32}\left\{I\left(\lambda,-\left(\frac{\pi}{2}+\alpha_1\right) ; \frac{\pi}{2}+\alpha_1+\alpha_2\right)+I\left(\lambda,-\left(\frac{\pi}{2}-\alpha_1\right) ; \frac{\pi}{2}-\alpha_1+\alpha_2\right)-\right.\\
        &\quad\quad-I\left(\lambda,-\left(\frac{\pi}{2}+\alpha_1\right) ; \pi+\alpha_1\right)-I\left(\lambda,-\left(\frac{\pi}{2}-\alpha_1\right) ; \pi-\alpha_1\right)-\\
        &\quad\quad-I\left(\lambda,-\left(\frac{\pi}{2}+\alpha_5\right) ; \pi+\alpha_5\right)-I\left(\lambda,-\left(\frac{\pi}{2}-\alpha_5\right) ; \pi-\alpha_5\right)+\\
        &\quad\quad\left.+I\left(\lambda,-\left(\frac{\pi}{2}+\alpha_5\right) ; \frac{\pi}{2}+\alpha_5+\alpha_4\right)+I\left(\lambda,-\left(\frac{\pi}{2}-\alpha_5\right) ; \frac{\pi}{2}-\alpha_5+\alpha_4\right)\right\}
        \end{aligned}
\end{equation}
where $I$ is the weight 3 polylogarithm function defined by
\begin{equation}\label{eq:defi}
    \begin{aligned}
        I(a,b;x) &:= \frac{1}{4 i} \int_{\frac{\pi}{2}}^{x} \left( \text{Li}_2(e^{2 i y}) - \text{Li}_{2}(e^{-2 i y}) \right) \, d\arctan(a \tan(b+y)) \\
        &= \frac{1}{8} \left\lbrack \text{Li}_{2,1} \left( \frac{a-1}{a+1} e^{-2ib}, \frac{a+1}{a-1} e^{2i (b+x)} \right) - \text{Li}_{2,1} \left( \frac{a-1}{a+1} e^{-2ib}, -\frac{a+1}{a-1} e^{2i b} \right) \right. \\
        &\quad - \left. \text{Li}_{2,1} \left( \frac{a+1}{a-1} e^{-2ib}, \frac{a-1}{a+1} e^{2i (b+x)} \right) + \text{Li}_{2,1} \left( \frac{a+1}{a-1} e^{-2ib}, -\frac{a-1}{a+1} e^{2i b} \right) \right. \\
        &\quad + \left. \text{Li}_{2,1} \left( \frac{a-1}{a+1} e^{2ib}, \frac{a+1}{a-1} e^{-2i (b+x)} \right) - \text{Li}_{2,1} \left( \frac{a-1}{a+1} e^{2ib}, -\frac{a+1}{a-1} e^{-2i b} \right) \right. \\
        &\quad - \left. \text{Li}_{2,1} \left( \frac{a+1}{a-1} e^{2ib}, \frac{a-1}{a+1} e^{-2i (b+x)} \right) + \text{Li}_{2,1} \left( \frac{a+1}{a-1} e^{2ib}, -\frac{a-1}{a+1} e^{-2i b} \right) \right\rbrack.
    \end{aligned}
\end{equation}
To express Kellerhals's formula~(\ref{eq:voldoubort}) in terms of the Gram matrix $Q$ of the full simplex, we can first write the dihedral angles in terms of the cross-ratios and then use~(\ref{eq:simp2moduli}). According to the hyperbolic geometry identities reviewed in Appendix~\ref{apdx:geom}, the map from moduli space to the dihedral angle can be expressed as
\begin{equation}\label{eq:moduli2angle5}
    \begin{aligned}
        & \alpha_i = \arccos \sqrt{ \frac{ z_{i-1,i} z_{i+1,i+2} }{ z_{i-1,i+1} z_{i,i+2} }} \,, \quad 1 \leq i \leq n-1 \,, \\
        & \alpha_0 = \arctan \sqrt{ - \frac{ z_{0,1} z_{2,5} z_{6,7} }{ z_{1,2} z_{5,6} z_{0,7} }} \,,  \\
        & \Theta = \arctan \sqrt{ - \frac{ z_{1,2} z_{3,4} z_{5,6} z_{0,7} }{ z_{0,1} z_{2,3} z_{4,5} z_{6,7} } } \,.
    \end{aligned}
\end{equation}
With all ingredients in place, we conclude that the volume of an ideal simplex in hyperbolic 5-space, (the type corresponding to the hexagon integral in six dimensions with massless internal propagators) is
\begin{equation}
\boxed{
    \text{Vol}(Q_{n=6}^{\text{ideal}}) =\sum_{\sigma \in S_{n-1}} \frac{ \text{sgn}(\sigma)\, \text{sgn}(\det \calQ^{\sigma}) }{\text{sgn}(\det Q)}  (\text{eq.~\ref{eq:voldoubort}}) \bigg|_{\text{eq.~(\ref{eq:moduli2angle5})}} \, \bigg|_{\text{eq.~(\ref{eq:simp2moduli}})}.
    }
\end{equation}
In practice this formula is significantly simpler than~(\ref{eq:volsimpsgn}) when $Q$ is ideal.

\acknowledgments

We are grateful to Jacopo Chen for pointing out an extra factor of 1/2, to Oliver Schlotterer for discussions on the single-valued map, and to Daniil Rudenko for encouraging comments. This work was supported in part by the US Department of Energy under contract DE-SC0010010 (Task F) and by Simons Investigator Award \#376208. CV was supported in part by grant 00025445 from Villum Fonden.

\appendix

\section{Some hyperbolic geometry}
\label{apdx:geom}

Since an orthoscheme is defined by a sequence of planes $(H_1,\ldots,H_n)$ such that $H_i$ and $H_j$ are orthogonal if $|i-j|\ge 2$, only the dihedral angle $\alpha_i$ between adjacent planes $H_i$ and $H_{i+1}$ can be nontrivial. This angle is given by
\begin{equation}
    \cos^2 \alpha_i = \frac{ ({ \calQ^{-1}_{i,i+1} })^2 }{ \calQ^{-1}_{i,i} \calQ^{-1}_{i+1,i+1} }\,,\qquad 1 \leq i \leq n-1 \,,
\end{equation}
where $\calQ^{-1}$ denotes the inverse of the matrix $\calQ$. Writing these out explicitly in terms of the entries of $\calQ$ and using~(\ref{eq:orthos2moduli}) one can also express the dihedral angles as
\begin{equation}\label{eq:moduli2angle}
    \cos^2 \alpha_i = \frac{z_{i-1,i} z_{i+1,i+2}}{z_{i-1,i+1} z_{i,i+2}} \,, \qquad 1 \leq i \leq n-1 \,.
\end{equation}

The relation~(\ref{eq:moduli2angle}) generalizes to cross-ratios of four arbitrary points as follows. For four vertices $v_a, v_b, v_c, v_d$ satisfying $1 \leq b \leq c \leq d$, consider the subscheme $\langle a,b,c,d \rangle$. Denote by $\langle v_b,\, v_c \rangle_{\langle a,b,c,d \rangle}$ the dihedral angle between the planes of $\langle a,b,c,d \rangle$ which are opposite to vertices $v_b$ and $v_c$, and similarly for $\langle v_a,\, v_b \rangle_{\langle a,b,c,d \rangle}$ and $\langle v_c,\, v_d \rangle_{\langle a,b,c,d \rangle}$. (Note that in this manner of notation $\langle v_a,\, v_c \rangle_{\langle a,b,c,d \rangle} = \langle v_b,\, v_d \rangle_{\langle a,b,c,d \rangle} = \langle v_a,\, v_d \rangle_{\langle a,b,c,d \rangle} = 0$.) Then we have the relations
\begin{equation}\label{eq:moduli2angleg}
    \begin{aligned}
        & \cos^2 \left( \langle v_b,\, v_c \rangle_{\langle a,b,c,d \rangle} \right) = \frac{z_{a,b} z_{c,d}}{z_{a,c} z_{b,d}} \,, \\
        & \cos^2 \left( \langle v_c ,\, v_d \rangle_{\langle a,b,c,d \rangle} \right) = \frac{z_{b,c} z_{d,n+1}}{z_{b,c} z_{c,n+1}} \,, && 1 \le a \le b-1\,,\\
        & \cos^2 \left( \langle v_a,\, v_b \rangle_{\langle a,b,c,d \rangle} \right) = \frac{z_{0,a} z_{b,c}}{z_{0,b} z_{a,c}} \,, && c+1 \le d \le n\,.
    \end{aligned}
\end{equation}
We see that cross-ratios of the configuration of the $z_i$ in $\mathcal{M}_{0,n+2}$ can be interpreted as (squares of) cosines of dihedral angles of the orthoscheme. The only exception is when $z_0$ and $z_{n+1}$ are both among the four points, in which case we have the special cross-ratio
\begin{equation}\label{eq:moduli2edge}
    \cosh^2 l_{a,b} = \frac{z_{0,a} z_{b,n+1}}{z_{0,b} z_{a,n+1}}\,, \qquad 1 \leq a<b \leq n
\end{equation}
where $l_{a,b}$ is the hyperbolic length of the edge between $v_a$ and $v_b$.

One can generalize the relations~(\ref{eq:moduli2angleg}) even further. For any subscheme $\langle I \rangle$ (defined in Sec.~\ref{sec:dissection}), where $I = \{ \rho_1, \rho_2,\ldots, \rho_m \}$ is ordered and $3 \leq m \leq n$, we have
\begin{equation}\label{eq:moduli2anglegg}
    \cos^2 \left( \langle v_{\rho_{i}} ,\, v_{\rho_{i+1}} \rangle|_{\langle I \rangle} \right) = \frac{z_{\rho_{i-1},\rho_i} \, z_{\rho_{i+1},\rho_{i+2}}}{z_{\rho_{i-1},\rho_{i+1}} \, z_{\rho_{i},\rho_{i+2}}}\,, \qquad 1 \leq i \leq m-1 \,,
\end{equation}
with boundary cases $\rho_0 := 0$ and $\rho_{m+1} := n+1$.

\section{Details on Rudenko's volume formula}
\label{sec:rudenkodetails}

In this section we review the key ingredients from~\cite{Rudenko} that are needed to write down the explicit formula for the volume of an odd-dimensional orthoscheme. The final result is given in~(\ref{eq:volinorthos}) but we begin with several key definitions.

A \emph{weighted alphabet} is a set of \emph{weighted letters} $\lbrack \varphi, m \rbrack$ where $\varphi$ is an arbitrary function and $m$ is a positive integer, called its weight. In the following few paragraphs we define an algebra on weighted alphabets.

We can use \emph{concatenation} of letters to define words which we denote by any one of $\lbrack \varphi_1, m_1 \rbrack \lbrack \varphi_1, m_1 \rbrack$, $\lbrack \varphi_1, m_1 \rbrack \otimes \lbrack \varphi_1, m_1 \rbrack$, or $\lbrack \varphi_1, m_1 | \varphi_1, m_1 \rbrack$ without distinction. The concatenation of a sequence of weighted alphabets is called a \emph{word}. The empty word is denoted by $1$.

The \emph{product} of two weighted letters is defined by
\begin{equation}
    \lbrack \varphi_1, m_1 \rbrack \cdot \lbrack \varphi_1, m_1 \rbrack := \lbrack \varphi_1 \varphi_2 , m_1+m_2 \rbrack \,.
\end{equation}
The product of a weighted letter and a word is defined by $\lbrack \varphi, m \rbrack \cdot 1 := 0$ and
\begin{equation}
    \lbrack \varphi_1, m_1 \rbrack \cdot \lbrack \varphi_2, m_2 | \cdots | \varphi_k, m_k \rbrack := \lbrack \varphi_1 \varphi_2, m_1 + m_2 | \cdots | \varphi_k, m_k \rbrack\,,
\end{equation}
and we extend these definitions to the vector space of (formal) linear combinations of words in the obvious way.

The \emph{quasi-shuffle product}, or the \emph{star product} of two words is defined by
\begin{equation}
    \lbrack \varphi, m \rbrack \star 1 = 1 \star \lbrack \varphi, m \rbrack := \lbrack \varphi, m \rbrack
\end{equation}
and
\begin{multline}
    (\lbrack \varphi_1, m_1 \rbrack \otimes \omega_1) \star (\lbrack \varphi_2, m_2 \rbrack \otimes \omega_2)
    := \lbrack \varphi_1, m_1 \rbrack \otimes ( (\omega_1) \star (\lbrack \varphi_2, m_2 \rbrack \otimes \omega_2) )\\ + \lbrack \varphi_2, m_2 \rbrack \otimes ( (\lbrack \varphi_1, m_1 \rbrack \otimes \omega_1) \star (\omega_2) ) + (\lbrack \varphi_1, m_1 \rbrack \cdot \lbrack \varphi_2, m_2 \rbrack) \otimes (\omega_1 \star \omega_2)\,,
\end{multline}
where $\omega_1$ and $\omega_2$ denote words of arbitrary length.

The cross-ratio of four points is denoted
\begin{equation}
    [z_0,\, z_1,\, z_2,\, z_3] := \frac{z_{01} z_{23}}{z_{03}z_{21}} \,.
\end{equation}
More generally, consider any even number of points $z = (z_0,z_1,\ldots,z_{n+1})$ in $\mathbb{P}^1$. For any (ordered) even subset of the points $z_P = \{z_{p_0}, z_{p_1} \ldots, z_{p_{2k+1}}\}$ with $p_0 < p_1 < \cdots < p_{2k+1}$, the generalized cross-ratio is defined as
\begin{equation}
    \gcr(z_P) := \left\{
        \begin{aligned}
            & \prod_{i=1}^k [z_{p_0} ,\, z_{p_{2i-1}} ,\, z_{p_{2i}} ,\, z_{p_{2i+1}}] && \text{if $p_0$ is even}, \\
            & \prod_{i=1}^{k} [z_{p_0} ,\, z_{p_{2i-1}} ,\, z_{p_{2i}} ,\, z_{p_{2i+1}}]^{-1} && \text{if $p_0$ is odd}.
        \end{aligned}
    \right.
\end{equation}

One can recursively define a map $T$ (called \emph{arborification}) from $z$ and any even subset $\{p_0,\ldots,p_{2k+1}\}$ of the ordered set $\{0,1,\ldots,n+1\}$ to the set of words by
\begin{equation}\label{eq:deftarb1}
    T_{P}(z) = \left\{
        \begin{aligned}
            & \sum_{\substack{0<i<j<2n+1 \\ i \text{ is odd} \\ j \text{ is even}}} T_{(0,i,j,2n+1)} \otimes \left( T_{(0 \cdots i)} \star T_{(i \cdots j)} \star T_{(j \cdots 2n+1)} \right) && \text{if $p_0$ is even,} \\
            & \sum_{\substack{1<i<j<2n+2 \\ i \text{ is even} \\ j \text{ is odd}}} \Big\{ T_{(1,i,j,2n+2)} \otimes \left( T_{(1 \cdots i)} \star T_{(i \cdots j)} \star T_{(j \cdots 2n+2)} \right) \\
            & \qquad\qquad\  +\, T_{(1,i,j,2n+2)} \cdot \left( T_{(1 \cdots i)} \star T_{(i \cdots j)} \star T_{(j \cdots 2n+2)} \right) \Big\} &&  \text{if $p_0$ is odd,} \\
        \end{aligned}
    \right.
\end{equation}
where
\begin{equation}\label{eq:deftarb2}
    \begin{aligned}
        & T_{(p_0, p_1, p_2, p_3)}(z) = \left\{
            \begin{aligned}
                & -\left\lbrack \frac{z_{p_0,p_1} z_{p_2,p_3}}{z_{p_0,p_3} z_{p_2,p_1}}, 1 \right\rbrack && \text{if $p_0$ is even,} \\
                & \left\lbrack \frac{z_{p_0,p_3} z_{p_2,p_1}}{z_{p_0,p_1} z_{p_2,p_3}}, 1 \right\rbrack && \text{if $p_0$ is odd}
            \end{aligned}
        \right. \\
         & T_{(p_0, p_1)}(z) = 1 \,.
    \end{aligned}
\end{equation}
As an example, consider $P = (0,1,2,3,4,5)$, which gives
\begin{equation}\label{eq:deftarb3}
    T_{P}(x) = \left\lbrack \frac{z_{01} z_{25}}{z_{05} z_{21}}, 1 \right. \left| \frac{z_{23} z_{45}}{z_{25} z_{43}}, 1 \right\rbrack  - \left\lbrack \frac{z_{01} z_{45}}{z_{05} z_{41}}, 1 \right. \left| \frac{z_{14} z_{32}}{z_{12} z_{34}}, 1 \right\rbrack + \left\lbrack \frac{z_{03} z_{45}}{z_{05} z_{43}}, 1 \right. \left| \frac{z_{01} z_{23}}{z_{03} z_{21}}, 1 \right\rbrack.
\end{equation}
On the other hand for $P = (1,2,3,4,5,6)$, we have
\begin{equation}
    \begin{aligned}
        T_{P}(z) &= \left\lbrack \frac{z_{16} z_{32}}{z_{12} z_{36}}, 1 \right. \left| \frac{z_{36} z_{54}}{z_{34} z_{56}}, 1 \right\rbrack  - \left\lbrack \frac{z_{16} z_{52}}{z_{12} z_{56}}, 1 \right. \left| \frac{z_{23} z_{45}}{z_{25} z_{43}}, 1 \right\rbrack\\
        &\qquad + \left\lbrack \frac{z_{16} z_{54}}{z_{14} z_{56}}, 1 \right. \left| \frac{z_{14} z_{32}}{z_{12} z_{34}}, 1 \right\rbrack
        + \left\lbrack \frac{z_{16}z_{23}z_{45}}{z_{12}z_{34}z_{56}}, 2 \right\rbrack.
    \end{aligned}
\end{equation}

The \emph{alternating multiple polylogarithm} associated to a word is defined by
\begin{equation}\label{eq:aliweightsymb}
    \text{ALi}(\lbrack \varphi_1, m_1 | \cdots | \varphi_k, m_k \rbrack) := \sum_{\epsilon_1, \ldots, \epsilon_k \in \{-1,1\}} \left(\prod_{i=1}^{k} \frac{\epsilon_i}{2}\right) \text{Li}_{m_1, \ldots, m_k} (\epsilon_1 \sqrt{\varphi_1}, \ldots, \epsilon_k \sqrt{\varphi_k})
\end{equation}
in terms of the familiar multiple polylogarithms (specifically, using the conventions of, for example, \cite{Duhr:2019tlz}). This definition extends to linear combinations of words (with coefficients $\pm 1$, which is all that is needed) by
\begin{equation}
    \text{ALi}(\omega_1 \pm \omega_2) = \text{ALi}(\omega_1) \pm \text{ALi}(\omega_2) \,.
\end{equation}

With all that preparation, we are now ready to present Rudenko's formula for the volume of the $(n{-}1)$-dimensional hyperbolic orthoscheme ${\cal{Q}}$ corresponding to the configuration $z = (z_0,\ldots,z_{n+1})$ of $n{+}2$ points in $\mathbb{P}^1$:
\begin{equation}
    \label{eq:volinorthos}
    \text{Vol}({\cal{Q}}) = \frac{1}{2^{\frac{n}{2}-1} \Gamma(\frac{n}{2})}\text{per} \left( \text{ALi}\big(T_{(0,\ldots, n+1)} (z) \big) \right),
\end{equation}
where $\text{per}$ is the real period defined in~\cite{Goncharov} and reviewed in the next appendix.

\section{Real periods}
\label{app:realperiod}

In this section we give some examples for the real period map, which was first introduced in~\cite{Goncharov}. The definition of the real period is\footnote{Here we use a definition different from that in~\cite{Goncharov} and that used in~\cite{Rudenko} by a factor of $i^{m+1}$, where $m$ denotes the weight of the function. This is in order to make the volume in~(\ref{eq:volinorthos}) always be a positive real number (for orthoschemes corresponding to Euclidean kinematics).}
\begin{equation}\label{eq:realperiod}
    \text{per}(G) := (2\pi)^{w(G)} \sum_{k=1}^{w(G)} \sum_{a_1 + \cdots + a_k = w(G)} (-1)^{k-1}\, \nabla ( \Delta_{a_1,\cdots, a_k} (G) ) \,,
\end{equation}
where $w$ denotes the weight, $\Delta$ is the coproduct (with $\Delta_{w(G)} (G) = G$), and $\nabla$ is defined as
\begin{equation}
    \nabla \left( \bigotimes_{i=1}^m f_i \right) :=
    \re\left( \prod_{i=1}^{m-1} \frac{f_i}{(2\pi i)^{w(f_i)}} \right)\, \im\left( \frac{f_m}{(2\pi i)^{w(f_m)}} \right) ,
\end{equation}
where we take $\im (x + i y) = i y$. Note that despite the name, the real period is only real-valued for functions of odd weight; for even weight it is purely imaginary.

The real periods are closely related to the single-valued map~\cite{Brown} more widely known in the physics literature (see for example~\cite{DelDuca:2016lad,Britto:2021prf}). Specifically, they are related by:
\begin{equation}
    \label{eq:per-sv-relation}
    \per(G(\vec{a};z)) =
    \begin{cases}
        -\frac{i}{2} \, \re\, \text{sv}(G(\vec{a};z)), \qquad & w(G)\text{ odd, }\\
        -\frac{i}{2} \, \im\, \text{sv}(G(\vec{a};z)), \qquad & w(G) \text{ even. }
    \end{cases}
\end{equation}

\paragraph{Weight 1}
The only weight 1 function is the logarithm $\log(x)$. It has only one non-vanishing coproduct component, $\Delta_1(\log(x)) = \log(x)$. Therefore, its real period reads
\begin{equation}
    \per(\log(x)) = -2\pi \nabla(\log(x)) = 2\pi\, \im\left( \frac{\log(x)}{2\pi i} \right) = -\re (\log(x)) = -\log(|x|)\,.
\end{equation}

\paragraph*{Weight 2}
Weight 2 functions are building blocks for the volumes of hyperbolic 3-simplices, and the box integral in 4 dimensions. We start with $\Li_2(x)$, whose coproduct components are
\begin{equation}
    \Delta_2 (\Li_2(x)) = \Li_2(x)\,, \qquad \Delta_{1,1} (\Li_2(x)) = \Li_1(x) \otimes \log(x)\,.
\end{equation}
Therefore the real period of $\Li_2(x)$ is
\begin{equation}
    \label{eq:per2}
    \begin{aligned}
        \text{per} (\Li_2(x)) &= 4\pi^2 \, \Big\{ \nabla(\Li_2 (x)) - \nabla\left(\Li_1(x) \otimes \log(x)\right)\Big\} \\
        &= 4\pi^2 \, \left\{ \im\left( \frac{\Li_2(x)}{(2\pi i)^2} \right) - \re\left( \frac{\Li_1(x)}{2\pi i} \right)\, \im\left( \frac{\log(x)}{2\pi i} \right) \right\} \\
        &= -\im\,\Li_2(x) + \im\,\Li_1(x) \, \re\log(x)\\
    \end{aligned}
\end{equation}
which is ($(-1)$ times) the Bloch-Wigner function. Similarly it is easy to see that the real period of $\Li_{1,1}(x,y)$ is
\begin{multline}\label{eq:per11}
    \per(\Li_{1,1}(x,y)) = -  \im\,\Li_{1,1}(x,y) - \re\,\Li_1(1/x)\, \im\,\Li_1(x y) \\
    + \im\,\Li_1(y)\, \re\,\Li_1(x) + \re\,\Li_1(y)\, \im\,\Li_1(x y)\,.
\end{multline}

\paragraph*{Weight 3}
Here we tabulate the real periods for weight 3 (multiple) polylogarithms, relevant for the volumes of hyperbolic 5-simplices and the hexagon integral in 6 dimensions:
\begin{equation}
    \label{eq:per3}
    \per(\Li_{3}(x)) = \re\,\Li_3(x) + \left( \re\log (x) \right)^2\, \re\,\Li_1(x) - \re\log (x)\, \re\,\Li_2(x)\,, \qquad\qquad\;
\end{equation}
\begin{equation}
    \label{eq:per12}
    \begin{aligned}
        \per(&\!\Li_{1,2}(x,y)) =
       \re\,\text{Li}_{1,2}(x,y) +\re\log \left({1/y}\right) \, \re\,\text{Li}_{1,1}(x,y)\\
       &-\im\,\text{Li}_1\left({1/x}\right) \, \re\log \left({1/y}\right)\, \im\,\text{Li}_1(x y)
         + \im\,\text{Li}_1(y) \, \re\log \left({1/y}\right)\, \im\,\text{Li}_1(x y)\\
         &+\im\,\text{Li}_1\left({1/x}\right)\, \im\,\text{Li}_1(x y) \, \re\log \left({1/x y} \right)
         - \im\,\text{Li}_2\left({1/x} \right)\, \im\,\text{Li}_1(x y)\\
         &+ \im\,\text{Li}_2(y)\, \im\,\text{Li}_1(x y) -\re\,\text{Li}_1(x) \, \re\,\text{Li}_2(y)
        + \re\,\text{Li}_1\left({1/x} \right) \, \re\,\text{Li}_2(x y)\\
        &- 2\, \re\,\text{Li}_1(x) \, \re\,\text{Li}_1(y) \, \re\log \left({1/y} \right)
         + \re\,\text{Li}_1\left({1/x} \right) \, \re\log \left({1/y} \right) \, \re\,\text{Li}_1(x y)\\
         &+ \re\,\text{Li}_1\left({1/x} \right) \, \re\,\text{Li}_1(x y) \, \re\log \left({1/x y} \right)
        - \re\,\text{Li}_1(y) \, \re\log \left({1/y} \right) \, \re\,\text{Li}_1(x y)\,,
    \end{aligned}
\end{equation}
\begin{equation}\label{eq:per21}
    \begin{aligned}
        \per(&\!\Li_{2,1}(x,y))
        = \re\,\text{Li}_{2,1}(x,y) -\re\log \left({1/y} \right) \, \re\,\text{Li}_{1,1}(x,y)\\
        &+\re\log \left({1/x y} \right) \, \re\,\text{Li}_{1,1}(x,y)
         - \im\,\text{Li}_1(x)\, \im\,\text{Li}_1(y) \, \re\log \left({1/y} \right)\\
         &+ \im\,\text{Li}_1\left({1/x} \right) \, \re\log \left({1/y} \right)\, \im\,\text{Li}_1(x y)
         - \im\,\text{Li}_1(y) \, \re\log \left({1/y} \right)\, \im\,\text{Li}_1(x y)\\
         &+\im\,\text{Li}_1(x)\, \im\,\text{Li}_1(y) \, \re\log \left({1/x y} \right)
         + \im\,\text{Li}_1\left({1/x} \right)\, \im\,\text{Li}_1(x y)\, \re\log \left({1/x y} \right)\\
         &+\im\,\text{Li}_2(x)\, \im\,\text{Li}_1(y)
        -\im\,\text{Li}_2(y)\, \im\,\text{Li}_1(x y)
        + \im\,\text{Li}_1(x y)\,  \im\,\text{Li}_2\left({1/x}\right)\\
        &-\re\,\text{Li}_1(y) \, \re\,\text{Li}_2(x y)
        + \re\,\text{Li}_1(x) \, \re\,\text{Li}_1(y) \, \re\log \left({1/y} \right) \\
        & - \re\,\text{Li}_1(x) \, \re\,\text{Li}_1(y) \, \re\log \left({1/x y} \right)
         - \re\,\text{Li}_1\left({1/x} \right) \, \re\log \left({1/y} \right) \, \re\,\text{Li}_1(x y) \\
         &+ \re\,\text{Li}_1\left({1/x} \right) \, \re\,\text{Li}_1(x y) \, \re\log \left({1/x y} \right)
         + \re\,\text{Li}_1(y) \, \re\log \left({1/y} \right) \, \re\,\text{Li}_1(x y)\\
         &- 2\, \re\,\text{Li}_1(y) \, \re\,\text{Li}_1(x y) \, \re\log \left({1/x y} \right),
    \end{aligned}
\end{equation}
\begin{equation}\label{eq:per111}
    \begin{aligned}
        \per(&\!\Li_{1,1,1}(x,y,z))
        = \im\,\text{Li}_1(z) \, \im\,\text{Li}_{1,1}(x,y) -\im\,\text{Li}_{1,1}\left(y,{1/x y} \right) \, \im\,\text{Li}_1(x y z) \\
        &+\im\,\text{Li}_{1,1}(y,z) \, \im\,\text{Li}_1(x y z)
         + \re\,\text{Li}_1\left({1/y} \right) \, \re\,\text{Li}_{1,1}(x,y z)\\
         &-\re\,\text{Li}_1(z) \, \re\,\text{Li}_{1,1}(x,y z) -\re\,\text{Li}_1(x) \, \re\,\text{Li}_{1,1}(y,z)\\
         &+ \re\,\text{Li}_1\left({1/x} \right) \, \re\,\text{Li}_{1,1}(x y,z) -\re\,\text{Li}_1(y) \, \re\,\text{Li}_{1,1}(x y,z) + \re\,\text{Li}_{1,1,1}(x,y,z) \\
         &+ \im\,\text{Li}_1(z) \, \re\,\text{Li}_1\left({1/x} \right) \, \im\,\text{Li}_1(x y)
         -\im\,\text{Li}_1(y) \, \im\,\text{Li}_1(z) \, \re\,\text{Li}_1(x) \\
         &- \im\,\text{Li}_1\left({1/x} \right) \, \re\,\text{Li}_1\left({1/y} \right) \, \im\,\text{Li}_1(x y z)\\
         &+\re\,\text{Li}_1\left({1/y} \right) \, \im\,\text{Li}_1(y z) \, \im\,\text{Li}_1(x y z)
         - \im\,\text{Li}_1(z) \, \re\,\text{Li}_1(y) \, \im\,\text{Li}_1(x y)\\
         &+\re\,\text{Li}_1(y) \, \im\,\text{Li}_1\left({1/x y} \right) \, \im\,\text{Li}_1(x y z)\
         - \im\,\text{Li}_1(z) \, \re\,\text{Li}_1(y) \, \im\,\text{Li}_1(x y z)\\
         &+\im\,\text{Li}_1\left({1/x} \right) \, \re\,\text{Li}_1\left({1/x y} \right) \, \im\,\text{Li}_1(x y z)
         - \re\,\text{Li}_1(z) \, \im\,\text{Li}_1(y z) \, \im\,\text{Li}_1(x y z)\\
         &+\re\,\text{Li}_1(x) \, \re\,\text{Li}_1(y) \, \re\,\text{Li}_1(z)
         -\re\,\text{Li}_1\left({1/x} \right) \, \re\,\text{Li}_1(z) \, \re\,\text{Li}_1(x y) \\
         &+ \re\,\text{Li}_1(y) \, \re\,\text{Li}_1(z) \, \re\,\text{Li}_1(x y)
         - 2\, \re\,\text{Li}_1(x) \, \re\,\text{Li}_1\left({1/y} \right) \, \re\,\text{Li}_1(y z)\\
         &+2\, \re\,\text{Li}_1(x) \, \re\,\text{Li}_1(z) \, \re\,\text{Li}_1(y z)
         + \re\,\text{Li}_1\left({1/x} \right) \, \re\,\text{Li}_1\left({1/y} \right) \, \re\,\text{Li}_1(x y z)\\
         &-\re\,\text{Li}_1(y) \, \re\,\text{Li}_1\left({1/x y} \right) \, \re\,\text{Li}_1(x y z)\\
        & + \re\,\text{Li}_1\left({1/x} \right) \, \re\,\text{Li}_1\left({1/x y} \right) \, \re\,\text{Li}_1(x y z)\\
         &-2\, \re\,\text{Li}_1\left({1/x} \right) \, \re\,\text{Li}_1(z) \, \re\,\text{Li}_1(x y z)
         + \re\,\text{Li}_1(y) \, \re\,\text{Li}_1(z) \, \re\,\text{Li}_1(x y z)\\
         &-\re\,\text{Li}_1\left({1/y} \right) \, \re\,\text{Li}_1(y z) \, \re\,\text{Li}_1(x y z)
         +\re\,\text{Li}_1(z) \, \re\,\text{Li}_1(y z) \, \re\,\text{Li}_1(x y z)\,.
    \end{aligned}
\end{equation}

\section{Explicit formula for the box integral in terms of alternating polylogs}
\label{sec:boxdetail}

In this section, we give the explicit formula for the volume of a hyperbolic 3-simplex, related to the box integral by~(\ref{eq:ivolrelation}). Combining~(\ref{eq:vol3ort}), (\ref{eq:simp2moduli}), and (\ref{eq:per11}), one obtains
\begin{equation}\label{eq:volq4}
    \begin{aligned}
        \text{Vol}(Q_{n=4}) = \frac{1}{2} & \Bigg\{\text{sgn}\left(\frac{(Q_{1,2} Q_{1,4}-Q_{1,1} Q_{2,4}) \bar{Q}_{3,4}}{(Q_{1,2} Q_{1,4}-Q_{2,2} Q_{1,4}-Q_{1,1} Q_{2,4}+Q_{1,2} Q_{2,4}) (\bar{Q}_{1,4}+\bar{Q}_{2,4}+\bar{Q}_{3,4})}\right) \\
        &\quad \times\per\left(\text{ALi}_{1,1}\left\lbrack 1-\frac{Q_{4,4} \bar{Q}_{4,4}}{\Delta^2},\frac{Q_{1,4}^2 \bar{Q}_{3,4}^2}{(Q_{1,2} Q_{1,4}-Q_{1,1} Q_{2,4}){}^2 (\Delta^2-Q_{4,4} \bar{Q}_{4,4})}\right\rbrack \right.\\
        &\qquad\qquad \left. + \text{ALi}_{1,1}\left\lbrack \frac{Q_{1,4}^2 (Q_{1,2}^2-Q_{1,1} Q_{2,2})}{(Q_{1,2} Q_{1,4}-Q_{1,1} Q_{2,4}){}^2},\frac{\bar{Q}_{3,4}^2}{\Delta^2 (Q_{1,2}^2-Q_{1,1} Q_{2,2})}\right\rbrack \right.\\
        &\qquad\qquad \left. - \text{ALi}_{1,1}\left\lbrack \frac{Q_{1,4}^2}{Q_{1,4}^2-Q_{1,1} Q_{4,4}},\frac{(Q_{1,4}^2-Q_{1,1} Q_{4,4}) \bar{Q}_{3,4}^2}{\Delta^2 (Q_{1,2} Q_{1,4}-Q_{1,1} Q_{2,4})^2} \right\rbrack \right) \Bigg\} \\
        & \qquad\qquad\qquad\qquad\quad\qquad\qquad\qquad\qquad\qquad\qquad\qquad + \text{(perm 1,2,3)} \,,\\
    \end{aligned}
\end{equation}
where
\begin{equation}
        \bar{Q}_{i,j} := (-1)^{i+j} \det Q[1 \cdots \hat{\imath} \cdots n; 1 \cdots \hat{\jmath} \cdots n]\,,\quad
        \Delta := \det Q \,.
\end{equation}
Now let us consider the massless limit, for which $Q_{i,i}\to 0$. In this limit, (\ref{eq:volq4}) simplifies to
\begin{equation}\label{eq:volq4massless}
    \begin{aligned}
        \text{Vol}(Q_{n=4}^{\rm ideal}) = \frac{1}{2} & \Bigg\{\text{sgn}\left(\frac{Q_{1,4}\bar{Q}_{3,4}}{(Q_{1,4} + Q_{2,4}) (\bar{Q}_{1,4}+\bar{Q}_{2,4}+\bar{Q}_{3,4})}\right) \per\left(\text{ALi}_{1,1}\left\lbrack 1,\frac{\bar{Q}_{3,4}^2}{Q_{1,2}^2\,\Delta}\right\rbrack \right) \Bigg\}\\
        & \qquad\qquad\qquad\qquad\quad\qquad\qquad\qquad\qquad\qquad\qquad\qquad {+} \text{(perm 1,2,3)} \,.\\
    \end{aligned}
\end{equation}
After using~(\ref{eq:ALidef}) and computing the real period of $\li_{1,1}$ via (\ref{eq:per11}) and using the identity
\begin{equation}\label{eq:11to2}
    \operatorname{Li}_{1,1}\left(a_1, a_2\right)=\mathrm{Li}_2\left(\frac{1-a_1}{1-a_2^{-1}}\right)-\mathrm{Li}_2\left(\frac{a_2}{a_2-1}\right)-\mathrm{Li}_2\left(a_1 a_2\right),
\end{equation}
one can check that~(\ref{eq:volq4massless}) finally evaluates to the familiar four-mass box function
\begin{equation}
    \frac{1}{i}\left[\text{Li}_2\left( \frac{w+v-u + \sqrt{\Delta}}{w+v-u - \sqrt{\Delta}} \right) - \text{Li}_2\left( \frac{w+v-u - \sqrt{\Delta}}{w+v-u + \sqrt{\Delta}} \right)\right] + \text{(cyclic $u, v, w$)}
\end{equation}
where
\begin{equation}
    u = Q_{1,2} Q_{3,4}\,, \quad v = Q_{1,4}Q_{2,3}\,, \quad w = Q_{1,3}Q_{2,4} \,, \quad \Delta = u^2+v^2+w^2-2(uv+uw+vw)\,.
\end{equation}

\end{document}